\documentclass[fleqn,usenatbib]{mnras}

\usepackage{newtxtext,newtxmath}
\usepackage[T1]{fontenc}
\usepackage{ae,aecompl}

\usepackage{graphicx}	% Including figure files
\usepackage{booktabs}
\usepackage{array}

\newcommand{\DM}[1]{$\text{DM}=#1$\,pc\,cm$^{-3}$}
\newcommand{\fnurl}[1]{\footnote{\url{#1}}}
\newcolumntype{C}{>{\begin{math}}c<{\end{math}}}%
\newcolumntype{L}{>{\begin{math}}l<{\end{math}}}%

\title[Single-pulse Searcher]{Single-pulse classifier for the\\LOFAR Tied-Array All-sky Survey}

\author[D. Michilli et al.]{%
D.~Michilli,$^{1,2}$\thanks{E-mail: danielemichilli@gmail.com}
J.~W.~T.~Hessels,$^{1,2}$
R.~J.~Lyon,$^{4}$
C.~M.~Tan,$^{4}$
C.~Bassa,$^{2}$ \newauthor
S.~Cooper,$^{4}$
V.~I.~Kondratiev,$^{2,3}$
S.~Sanidas,$^{4,1}$
B.~W.~Stappers,$^{4}$
J.~van~Leeuwen$^{2,1}$
\\
$^{1}$Anton Pannekoek Institute for Astronomy, University of Amsterdam, Science Park 904, 1098 XH Amsterdam, The Netherlands\\
$^{2}$ASTRON, the Netherlands Institute for Radio Astronomy, Postbus 2, 7990 AA, Dwingeloo, The Netherlands\\
$^{3}$Astro Space Centre, Lebedev Physical Institute, Russian Academy of Sciences, Profsoyuznaya Str. 84/32, Moscow 117997, Russia\\
$^{4}$Jodrell Bank Centre for Astrophysics, School of Physics and Astronomy, The University of Manchester, Manchester M13 9PL, UK\\
}

\date{Accepted XXX. Received YYY; in original form ZZZ}
\pubyear{2017}

\begin{document}
\label{firstpage}
\pagerange{\pageref{firstpage}--\pageref{lastpage}}
\maketitle

\begin{abstract}
Searches for millisecond-duration, dispersed single pulses have become a standard tool used during radio pulsar surveys in the last decade. 
They have enabled the discovery of two new classes of sources: rotating radio transients and fast radio bursts.
However, we are now in a regime where the sensitivity to single pulses in radio surveys is often limited more by the strong background of radio frequency interference (RFI, which can greatly increase the false-positive rate) than by the sensitivity of the telescope itself.
To mitigate this problem, we introduce the Single-pulse Searcher (\textsc{SpS}).
This is a new machine-learning classifier designed to identify astrophysical signals in a strong RFI environment, and optimized to process the large data volumes produced by the new generation of aperture array telescopes.
It has been specifically developed for the LOFAR Tied-Array All-Sky Survey (LOTAAS), an ongoing survey for pulsars and fast radio transients in the northern hemisphere.
During its development, \textsc{SpS} discovered 7 new pulsars and blindly identified $\sim$80 known sources.
The modular design of the software offers the possibility to easily adapt it to other studies with different instruments and characteristics. 
Indeed, \textsc{SpS} has already been used in other projects, e.g.\ to identify pulses from the fast radio burst source FRB~121102.
The software development is complete and \textsc{SpS} is now being used to re-process all LOTAAS data collected to date.
\end{abstract}

% Select between one and six entries from the list of approved keywords.
% Don't make up new ones.

\section{Introduction}\label{sec:introduction}
% Justification of low-frequency observations
The first pulsar was discovered by recording its single pulses at $\sim80$\,MHz using an aperture array \citep{Hew68}.
In later studies, folding and Fourier-based techniques were used to take advantage of the pulsar periodicity.  Many pulsar observations shifted to higher observing frequencies around $1.4$\,GHz, where the separation between pulsar signals and sky brightness is maximum for most of the pulsar population \citep{Cli86}.  Furthermore, phased arrays were generally replaced by large single dishes, which remove the complexity of signal correlation permitting an increase in telescope sensitivity and bandwidth \citep{Gar12}.
However, in recent years the increase in available computing power makes it possible to build phased aperture array telescopes with sensitivities and bandwidths that outperform traditional single dishes at low radio frequencies, offering a larger field-of-view (FoV) and more flexible instruments \citep[][]{Haa13,Tay12,Tin13}.
This enables an exploration of a parameter space that is complementary to other searches: e.g., it is possible to detect sources having a spectrum steeper than the sky background \citep[spectral index $\alpha \sim -2.5$,][]{Moz17}, which are likely too faint to be detected at higher frequencies.
In addition, the larger FoV improves survey speed, and makes all-sky searches tractable.

% LOTAAS survey definition
The most sensitive phased aperture array telescope to date is the LOw Frequency ARray \citep[LOFAR,][]{Haa13,Sta11}. 
For example, its large collecting area already enabled detailed studies of archetypal sources, and the properties of the known pulsar population. 
The former is exemplified by the discovery of radio/X-ray mode switching in PSR~B0943+10 \citep{Her13}. 
Examples of the latter are the pulsar census results presented by \citet{Kon16} and \citet{Bil16}.
Beyond these known sources, many pulsars in our Galaxy remain undetected. 
It was recognised early on that LOFAR has great potential for discovering these \citep{Lee10}. 
Pilot surveys placed limits on the occurrence of fast transients at low frequencies, and discovered the first two radio pulsars with LOFAR \citep{Coe14}. 
We are now performing a full, sensitive survey of the northern hemisphere called the
LOFAR Tied-Array All-Sky Survey \citep[LOTAAS,][]{Coe14,San18}\fnurl{http://www.astron.nl/lotaas}. 
LOTAAS has already demonstrated its ability to find new pulsars using periodicity searches (with over 80 discoveries to date) and in this paper we focus on discoveries made through single pulses.
The long dwell time (1\,hr) and large FoV ($\sim 10$\,sq.\,deg. per pointing) of LOTAAS also make the survey potentially well placed to discover fast radio transients, as long as they are not strongly affected by scattering or dispersive smearing.

\subsection{Signal classification}
% Effect of RFI
An increasing issue for pulsar surveys is the presence of radio-frequency interference (RFI) produced by several devices, which can mimic the behaviour of astrophysical signals and limit survey sensitivities \citep[e.g.][]{Lyo16}.
The large number of RFI detections makes it impractical to visually inspect and follow-up all the detected signals.
This is a worsening problem caused by the increasing number of devices emitting radio waves and the improvements in telescope characteristics, such as sensitivity, dwell time and bandwidth.
Therefore, this is a major challenge for next-generation radio telescopes and in particular the Square Kilometre Array \citep[SKA,][]{Ell04}.
An improvement is obtained by building telescopes in RFI-free zones, areas where the human presence is minimal.
However, the radio emissions of air planes and satellites are still present.

% Classifiers
In order to lower the number of detections to be inspected by eye, many automated classifiers have been developed for pulsar surveys (see \citealt{Lyo16} and references therein for a summary).
These automatic classifiers evolved from simple heuristics and thresholds on S/N \citep[e.g.][]{Cli86} to semi-automated ranking algorithms \citep[e.g.][]{Lee13}.
Also the graphical representation of the detected signals evolved, visualizing  increasing information describing their parameters \citep[e.g.][]{Bur06}.
Machine learning (ML) techniques began to be used to evaluate heuristic performance, set threshold values and perform the classification \citep[e.g.][]{Eat10}.
Most recently, significant efforts are being spent in the selection of pulsar signals to deal with the large number of detections from new, sensitive radio telescopes \citep[e.g.][]{Yao16,For17,Bet17}.

Here a branch of ML called statistical classification \citep{Mit97} is used to filter to select astrophysical signals. This requires first acquiring a large set of candidate examples for which the ground truth origin or `class' is known.  When there are two classes under consideration, the classification task is termed `binary'. In binary problems the targets, i.e. astrophysical signals, belong to the positive class. The negative class describes all other examples (e.g.\ RFI or noise). In either case the examples must be characterised via one or more variables commonly known as `features'. Features are numerical or textual descriptors that summarise a candidate in some relevant way (e.g.\ S/N, pulse width, etc.). Features must be extracted by algorithms for each candidate, and linked to their true class `labels'. When combined, this information forms what is known as a `training set'.  Using supervised learning it is possible to `learn' a mathematical function from this training set, that can automatically perform a similar mapping on new data. This process is known as `training'. The training process aims to split the training data into their respective classes, by using the inherent differences in the feature distributions to separate them. The learning process is guided via a heuristic, most often error minimisation, that quantifies how many errors are made.  Each correct positive classification is known as a True Positive (TP), whilst an incorrect positive classification is known as a False Positive (FP). Similarly, negative classifications can be described in terms of True Negative (TN) and False Negative (FN) predictions. Together, these four outcomes form the so-called confusion matrix, used to assess how successfully an algorithm has learned the mapping. 

For a good classification, it is essential to have features that permit to separate the data into positive and negative classes.
Therefore, specific algorithms must be designed to extract such features.
Moreover, these algorithms need to be fast and efficient in order to keep the computing time low.
These algorithms can be designed by looking at the different properties of the members of the positive and negative classes.
The classification success is usually determined during a `testing' phase, conducted on an independent sample of candidate examples.  If the model learned during training performs well during testing (produces few FPs and FNs) it can be used to derive predictions for new previously unseen data. An algorithm will usually be successful if  trained on a large representative sample of data.
Different heuristics can be used to evaluate single features, such as the information gain \citep[also known as mutual information,][]{Bro12}, a measure of the correlation between a feature and the target variable \citep{Lyo16}.
Also, several metrics exist to evaluate the performance of the whole set of features in classifying the data.
In this study, we made use of standard metrics such as the false negative rate (FNR) and the false positive rate (FPR), which must be as low as possible for a good classification, the true positive rate (TPR, also known as recall), the positive predictive value (PPV, also known as precision), the accuracy (ACC), the G-Measure (G-M, the geometric mean of recall and precision) and the F-Measure (F-M, the harmonic mean of recall and precision), which must be as high as possible for a good classification \citep[e.g.][]{Pow11}.
All these metrics assume values between zero and one.

\subsection{Single-pulse searches}
% Justification of single pulse search
Soon after the initial discovery of pulsars, surveys began taking advantage of the inherent periodicity of pulsar signals to improve search sensitivity \citep{Lor04}. 
This is usually achieved by performing a Fourier transform of the time-series.
However, this technique greatly decreases the sensitivity to sources whose emission is not regular over time \citep{Mcl06}.
Therefore, it has become standard procedure to include single-pulse searches in pulsar surveys to avoid missing sources with large pulse amplitude variations or a large null fraction \citep{Lor04}. 
Moreover, two new classes of sources discovered in recent years have created new interest in single-pulse searches: rotating radio transients \citep[RRATs,][]{Mcl06} and fast radio bursts \citep[FRBs,][]{Lor07}.
The former class is composed of pulsars whose emission is so sporadic in time that they are missed by periodicity searches. 
Typical RRAT pulse rates range from as many as one per few seconds, up to one per several hours\fnurl{http://astro.phys.wvu.edu/rratalog}.
The FRB class \citep{Tho13} is composed of extra-galactic radio flashes \citep{Ten17}. 
To date, only one has been observed to repeat \citep{Spi16} and no periodicity has been detected \citep{Sch16}.

% Method of sp search
A typical search for single pulses is performed after de-dispersing the data collected by the telescope.
This aims to correct for the frequency-dependent delay induced by the interaction of the radio waves from the source with free electrons present along the line of sight.
The amount of dispersive delay exhibited by a signal is proportional to the dispersion measure (DM, which is the column density of free electrons). Since the DM of the source is unknown {\it a priori}, it is necessary to de-disperse the signal at many trial values.
The time-series resulting from the addition of all frequency channels is then searched for single pulses.
This is usually achieved by convolving each de-dispersed time-series with a top-hat function of variable width $W$.
The properties of the function and the convolution are recorded every time the resulting signal-to-noise ratio (S/N) is above a certain threshold (see \citealt{Lor04} for a detailed discussion).

\subsection{LOTAAS Single-pulse Searcher}
% RFI in LOTAAS
The number of signal detections arising from RFI is particularly large for LOTAAS because the LOFAR Core is in a  region of high population density.
In addition, the high sensitivity and long dwell time offered by the survey and the large parameter space necessary to be searched at these low frequencies increase the number of false detections.
For these reasons, an automated classifier has been developed to classify the periodic signals of LOTAAS \citep[][]{Lyo16,Tan18}.
It uses the advantages of statistical classification to build a solid statistical framework for rejecting RFI. 

% Single-pulse classifiers
The presence of RFI is particularly problematic for single-pulse searches.
In fact, as opposed to periodicity searches, it is not possible here to filter out aperiodic signals.
Usually, single-pulse classifiers take advantage of pulse shape in the frequency-time domain where, as opposed to RFI, astrophysical signals are expected to typically be broadband and dispersed \citep{Lor04}.
As opposed to periodicity searches discussed earlier, only a few classifiers have been reported to specifically sift single-pulse detections.
\citet{Kea10} subtracted the non-dispersed signal from their data \citep{Eat09} and used spatial information from a multi-beam survey to discover ten new RRATs.
\citet{Spi12} developed a multi-moment technique useful for quantifying the deviation present in the pulse intensity at different frequencies. 
\citet{Kar15} presented \textsc{RRATtrap}, which uses manually-set thresholds to discriminate astrophysical signals based on their S/N behaviour as a function of DM.
Similarly, \citet{Den16} developed \textsc{Clusterrank}, which rejects RFI instances that deviate from the theoretical relation between signal strength, width and DM  \citep{Cor03}.
\citet{Dev16} studied different ML algorithms applied to their single-pulse classifier.
For binary classification, a Random Forest (RF) trained on an oversampled set (RF$^2_\text{over}$) performed the best in their case.

% Aim of the paper
Here we present a new single-pulse classifier, Single-pulse Searcher \citep[\textsc{SpS},][]{mic18}, which is able to discriminate astrophysical signals from RFI with high speed and accuracy.
In its current implementation (LOTAAS Single-pulse Searcher, \textsc{L-SpS}), it has been specifically designed to process LOTAAS data.
LOTAAS observations and data processing are presented in \textsection\ref{sec:observations}; the \textsc{SpS} classifier is presented in \textsection\ref{sec:L-SpS}; first discoveries are presented in \textsection\ref{sec:results}; and conclusions and future developments are discussed in \textsection\ref{sec:conclusions}.

\section{Observations}\label{sec:observations}
% LOTAAS observations
LOFAR is a radio telescope composed of thousands of antennas, which are grouped into stations that are distributed across the Netherlands and other European countries \citep{Haa13,Sta11}.
LOTAAS observations \citep{Coe14} are performed using the $12$ sub-stations of the Superterp, a circular area with $\sim 350$-m diameter where the station concentration is highest.
The High-Band Antennas (HBAs) are used to observe between $119-151$\,MHz.
Signals from different stations are added both coherently (`tied-array' beams) and incoherently.
The beams are divided over three sub-array pointings (SAPs, which are pointing directions formed at station level).
A total of 222 simultaneous beams on sky are produced per pointing: $3$ incoherent beams, a hexagonal grid of $61\times3$ coherent beams and $12\times3$ additional coherent beams that are scattered around the central grid, and which can be pointed to known sources within the FoV.
The incoherent beams have a total FoV of $\sim30$\,deg$^2$, while the coherent beams have a  total FoV of $\sim10$\,deg$^2$ and a sensitivity $\sim \sqrt[]{12} = 3.5$ times higher than that of the incoherent beams.
Each LOTAAS pointing has a 1-hr dwell time and a time resolution of $0.492$\,ms. 
The whole survey is divided into three passes and, upon survey completion, each sky position will be covered three times by the incoherent beams and once by a coherent beam.

% Processing pipeline
Data are processed using a pipeline based on \textsc{PRESTO} \citep{Ran01}\footnote{\url{https://www.cv.nrao.edu/~sransom/presto/}} described in detail by \citet{San18}, which runs on the Dutch national supercomputer Cartesius\fnurl{https://userinfo.surfsara.nl/systems/cartesius}.
After using the \textsc{rfifind} algorithm to filter RFI in the time-frequency domain, the signal is incoherently de-dispersed and frequency channels are added together using \textsc{mpiprepsubband} in order to form time-series data.
Each time-series is searched for both periodic signals and single-pulse peaks.
The latter is performed using \textsc{single\_pulse\_search.py}, which convolves box-car functions of various lengths between $0.5$ and $100$\,ms.
Single-pulse peaks with a S/N higher than 5 are stored for further grouping and sifting.

A total of $\sim 10^4$ DM trials is performed between DM values of $0$~--~$550$\,pc\,cm$^{-3}$ with steps between DM trials of $0.01$~--~$0.1$\,pc\,cm$^{-3}$.
Given a total of $\sim 7 \times 10^6$ time samples per time-series, this implies a grid in the DM~--~time space with $7\times10^{10}$ pixels, each one potentially containing an astrophysical signal, for each of the $222$ beams in every pointing.
Considering only ideal random noise, this implies an expected number of spurious detections ($\geq 5$\,$\sigma$) for each observation of $\sim45$ \citep{Cor03}.
Instead, a typical observation produces $\sim10^7$-$10^8$ detections above $5 \sigma$, demonstrating the huge impact that RFI and non-stationary noise have on the single-pulse search.
For comparison, the number of detections generated above $6 \sigma$ is roughly an order of magnitude less and those above $7 \sigma$ are normally half of that. 
An automated algorithm capable of sifting these detections to identify the rare astrophysical signals is thus necessary.

\section{LOTAAS Single-pulse Searcher}\label{sec:L-SpS}
% SpS introduction
\textsc{L-SpS} \citep{mic18} is a new sifting algorithm designed for single-pulse searches in a strong-RFI environment, and specifically designed for the LOTAAS survey.
The software uses ML techniques to differentiate RFI from astrophysical signals, classify interesting signals and produce diagnostic plots.

\subsection{Algorithm operation}
The aim of the program is to produce diagnostic plots for the very best detections in a typical pulsar single-pulse search.
Three steps are necessary to achieve this result. 

\subsubsection{Events}
% Event
The script \textsc{single\_pulse\_search.py} included in \textsc{PRESTO} outputs information on each detected signal, specifically the DM, the time with respect to the beginning of the time-series and the selected width of the kernel function.
These values are stored for each signal as one line in a text file.
We define one of such lines as an \emph{event}.
The \textsc{PRESTO}-based pipeline is run for every LOTAAS pointing.
For each beam, \textsc{L-SpS} copies the events detected into the memory of the computer and stores them into a readily accessible HDF5 database\fnurl{https://www.hdfgroup.org}.

An impulsive signal having a large enough S/N, either RFI or astrophysical, will be detected in multiple time-series de-dispersed at nearby values. 
Therefore it will produce a number of events close in time and DM.
A broadband signal exhibits a decreasing S/N as the trial DM used moves further from the actual DM value because the pulse becomes increasingly smeared in time.
This smearing is not symmetric and advances or delays the events for DM values higher or lower than the actual value, respectively.
Therefore, the events from a broadband signal will lie on a line in the DM~--~time space whose slope is given by 
\begin{equation}\label{eq:delay}
\frac{\Delta t}{\Delta\text{DM}} = \frac{k}{2} \left( \nu_{m}^{-2} - \nu_{M}^{-2} \right),
\end{equation}
where $k=(4148.808\pm0.003)$\,MHz$^2$\,pc$^{-1}$\,cm$^3$\,s is the dispersion constant \citep{Lor04}.
This equation is half the value of the DM delay between the minimum ($\nu_{m}$) and maximum ($\nu_{M}$) observing frequencies.
In \textsc{L-SpS}, the slope defined by Eq.~\ref{eq:delay} is corrected so that events belonging to the same broadband signal are simultaneous.

\subsubsection{Pulses}\label{sec:pulses}
% Pulse
All events close in time and DM are grouped together to form a \emph{pulse}.
The grouping is performed by a friends-of-friends algorithm \citep[e.g.][]{Huc82,Pre82} designed to be highly efficient and able to process one million events in approximately half a second.
Such an algorithm starts from the first event in the list and labels the closest other event in the DM~--~time space to be part of the same pulse if they are within a certain range.
The thresholds on time and DM ranges between successive events were set empirically to $30$\,ms and $20$ DM trials (i.e. between $0.2$ and $2$\,pc\,cm$^{-3}$ depending on the DM trial step size), respectively.
The algorithm subsequently processes all the events with the same conditions.
The characteristics of each pulse (i.e. time of arrival, DM, width, S/N) are derived from the brightest event forming that pulse.
Pulses formed by less than five events are considered spurious or too weak for subsequent analysis and therefore removed.
Weak narrow-band signals, which produce pulses with constant S/N, are mitigated by removing pulses with $\text{S/N} < 6.5$.
Also, pulses with $\text{DM} < 3$\,pc\,cm$^{-3}$ are removed to avoid the contamination of broad-band RFI near \DM{0}.
Tests are ongoing to try to lower this threshold on the DM.
ML classification is then applied to the pulses in order to discriminate RFI from astrophysical signals, as discussed in \textsection\ref{sec:ML}.

After the pulses labelled as RFI by the ML classifier are removed, those positively classified are further filtered based on spatial information.
Given the sensitivity pattern of the LOTAAS beams, astrophysical signals will have a maximum S/N in the beam closest to the actual source position and decreasing S/N values in beams farther away. 
Given the complicated pattern of the side lobes, weaker detections can be expected in distant beams for bright signals.
For each pulse detected in a coherent beam, all the events in the other non-adjacent coherent beams of the same SAP are loaded if they are in a time window four times the pulse width and a DM window of $0.4$\,pc\,cm$^{-3}$ around the pulse.
Also, these events must have a S/N larger than half the pulse's S/N in order to only remove signals with a nearly constant strength over many beams.
If more than four additional beams contain events selected with those criteria, then the pulse is discarded.
The spatial comparison is computationally expensive because of the many events to load and select.
Therefore, it is implemented after the ML classifier has already removed a large fraction of the pulses.

\subsubsection{Candidates}\label{sec:candidates}
% Candidate
In every beam, positively classified pulses (occurring at different times) are grouped into \emph{candidates} according to their proximity in DM.
The maximum DM spread over which two pulses are considered to come from
the same source is empirically set to $0.3$\,pc\,cm$^{-3}$, corresponding to 3--30 DM trials.
Candidates in different beams within this DM range are considered to be produced by the same signal and only the brightest is retained.
The characteristics of each candidate are computed, i.e. beam number, average DM, cumulative S/N, number of pulses detected and time of arrival if only one pulse is present, otherwise the period is calculated using the \textsc{PRESTO} routine \textsc{rrat\_period}.
Candidates formed by a single pulse are considered only if they satisfy $\text{S/N} > 10$ since weaker candidates would not be visible in the dynamic spectrum, meaning that their astrophysical nature can not be verified.
For the same reason, candidates formed by multiple pulses must have a cumulative $\text{S/N} > 16$.

A maximum of ten bright candidates (i.e. candidates with the largest cumulative S/N) are processed per beam.
Typically, a lower number of candidates per beam is produced, as discussed in \textsection\ref{sec:performance}.
This limit is set empirically to avoid processing too many spurious candidates, particularly in beams polluted by RFI.
For the same reason, a processing limit is set by considering only the 50 brightest candidates per observation.
As indicated in \citet{Lyo16}, this is not an ideal procedure.
However, typical observations produce a factor of $\sim3$ less candidates.
Diagnostic plots are then generated for every selected candidate, as discussed in \textsection\ref{sec:L-SpS:plot}.

\subsection{Machine Learning classifier}\label{sec:ML}
% Astrophysical vs RFI pulse characteristics
Astrophysical signals of interest are different from RFI in that they are usually broadband, dispersed and localized on the sky, while normally the latter is either narrow-band or not dispersed and often detected in multiple beams because the source is local.
This implies that astrophysical signals produce pulses that peak in S/N in a certain beam and at a certain DM~$>0$\,pc\,cm$^{-3}$, while RFI will often have similar S/N in all the beams and constant S/N at different DM trials if narrow-band, or peaked at DM~$=0$\,pc\,cm$^{-3}$ if it is broadband.
However, the real-world situation is often more complicated because of statistical and non-stationary noise (which tends to mask these differences especially for weak signals) as well as artefacts introduced by the telescope.
Also, RFI can mimic a dispersive delay \citep[e.g.][]{Pet15}, for example when multiple signals occur simultaneously in the de-dispersed data.
An astrophysical signal can also be masked by simultaneous RFI.
In addition, both RFI and astrophysical signals can have complex structures that complicate their S/N curve as a function of DM.
Finally, some sources of interference like aeroplanes, radars and signals reflected by the ionosphere can be beamed and thus appear localized on the sky.

% Application of statistical classification
We make use of ML techniques in order to effectively differentiate RFI and astrophysical signals and to have a statistical foundation to evaluate the performance of our features.
The statistical algorithm chosen to build the model and perform the classification is the Gaussian-Hellinger Very Fast Decision Tree \citep{Lyo14}, a tree learning algorithm \citep{Mit97} based on the Very Fast Decision tree \citep[VFDT;][]{Hul01}.
The algorithm is designed to deal with imbalanced problems, where the target class, in our case astrophysical single-pulse events, are outnumbered by the instances we wish to reject \citep{Lyo16}.
The data mining tool WEKA\fnurl{http://www.cs.waikato.ac.nz/ml/weka} was used to run this algorithm in order to evaluate classification performance, to build a classification model and to perform the classification of new instances.
The other algorithms included in the standard \textsc{WEKA} distribution have been tested using the training set described in the next subsection but none clearly outperformed the VFDT.

\subsubsection{Building of the training set}\label{sec:training_set}
\begin{figure}
\centering
\includegraphics{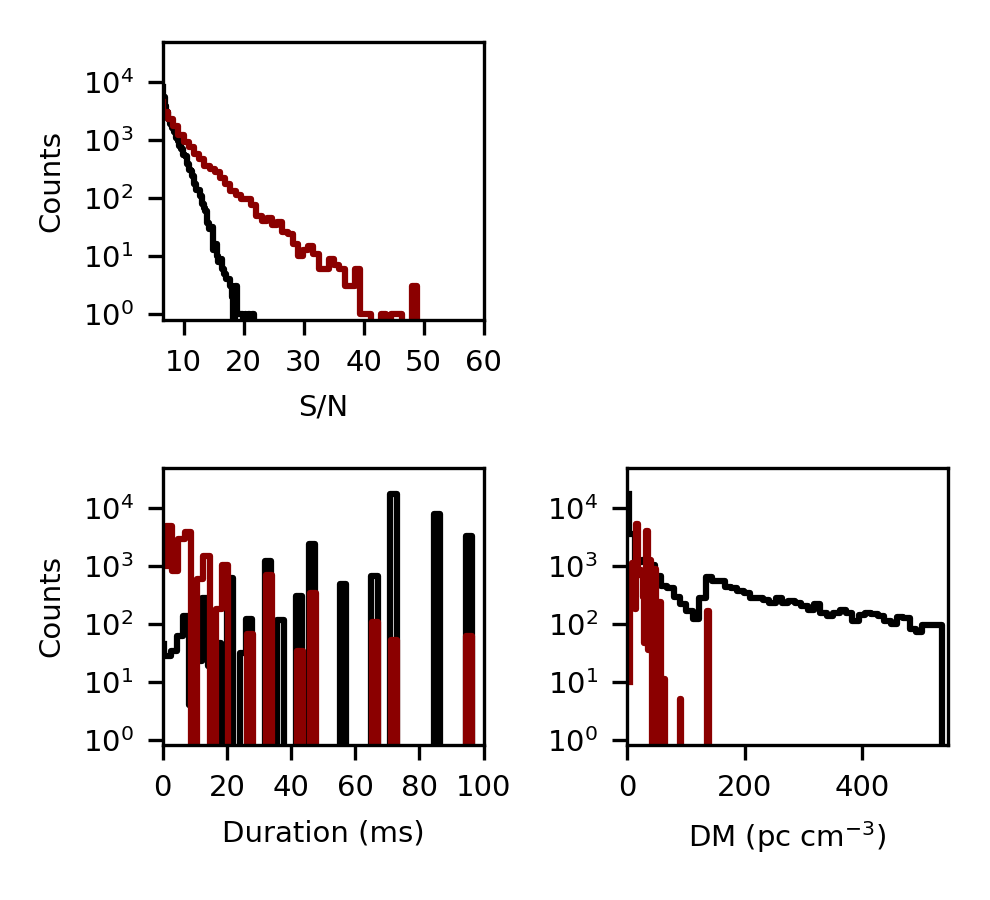}
\caption{Comparison of the duration, S/N and DM of the RFI (black lines) and astrophysical (red lines) instances used in the LOTAAS training set.
}
\label{img/set_distribution}
\end{figure}

% Training set
A set of manually-labelled instances was selected to evaluate the features used and to build the classification model.
A total of 35,063 instances of RFI were randomly chosen from various LOTAAS observations where no astrophysical sources could be identified by eye.
In addition, 18,003 pulses were chosen from $47$ known pulsars selected in an equal number of beams.
All the pulses in these beams were included in the training set if they were detected at the pulsar DM and were within $0.1\%$ of the expected time of arrival given the pulsar period.
The same conditions applied to random DM and period values, in beams without any known pulsar, usually yielded no pulses to be selected.
Given the costs associated with labelling data, the possibility of human error, and because the ground truth labels can only be confirmed via re-observing, it is possible for some training examples to be incorrectly labelled. Though this is unlikely, a small proportion of training samples may be affected in this way. Such mislabelled examples, whilst undesirable, have little impact on our results. This is because we used a sufficiently large collection of labelled training samples describing all classes, which compensates. In any case, such mislabelled examples can often help prevent a classifier from over-fitting, which reduces real-world performance \citep{Dud00,Bis06}.
Using this selection process, also pulses affected by noise or simultaneous RFI, which are desirable to be retained, were included.
In this way, we aimed to reduce potential classification bias against the rare class.
The distribution of some parameters of RFI and astrophysical instances in the training set is shown in Fig.~\ref{img/set_distribution}.

\subsubsection{Feature design}
% Filters generation
We developed different algorithms to extract features used to select relevant signals.
These algorithms were created by analysing individual characteristics of RFI instances.
At the end of the process, individual features had been evaluated according to the value of their information gain.
Features with a low information gain value were removed until a peak in the number of correctly classified instances was reached.
Residual redundant features were subsequently removed via an iterative method, manually deleting all the features one by one and calculating the number of correctly classified instances for each configuration, until a peak was reached.
\citet{Den16} developed a classification algorithm based on Eq.~12 of \citet{Cor03} that they reported performing well.
However, we could not use the same algorithm to extract new features for \textsc{L-SpS} because it was too slow in our implementation and required a large number of events per pulse.

% Selected filters
At the end of this process, the following five features were found to yield the highest number of correctly classified instances.
They are sorted in descending order according to their information gain (Table~\ref{tab:filters}):
\begin{enumerate}
 \item\label{feat1} Pulse width, W. 
 \item\label{feat2} Weighted mean of the pulse DM.
 \item\label{feat3} Excess kurtosis of the width curve as a function of DM. This curve is represented in red in Fig.~\ref{img/diagnostic_plot}h. 
 \item\label{feat4} Excess kurtosis of the S/N curve as a function of DM. This curve is represented in black in Fig.~\ref{img/diagnostic_plot}h.
 \item\label{feat5} Pulse S/N.
\end{enumerate}
Features \ref{feat2}, \ref{feat3} and \ref{feat4} have been adapted from \citet{Tan18}.

\begin{table}
\centering
\caption{ML features selected for \textsc{L-SpS} to separate astrophysical pulses from RFI and their value of information gain.
The subscript e indicates a single event within the pulse and $\sigma$ is the standard deviation. The rest of the symbols are defined in the text.}
\label{tab:filters}
{\renewcommand{\arraystretch}{2.}%
\begin{tabular}{lCC}
\toprule
  & \text{Feature} & \text{Inf. gain} \\
\midrule
\ref{feat1} & \text{W} = \text{W}_\text{e}(\max(\text{S/N}_\text{e})) 	& 0.74 \\
\ref{feat2} & \overline{\text{DM}} = \frac{\sum_\text{e} \text{DM}_\text{e}\,\text{S/N}_\text{e}}{\sum_\text{e}\text{DM}_\text{e}}  	& 0.71\\
\ref{feat3} & \text{k}_\text{W} = \frac{\sum_\text{e} (\text{DM}_\text{e}-\overline{\text{DM}})^4\text{W}_\text{e}}{\sigma^4(\text{W}_\text{e})\sum_\text{e}\text{W}_\text{e}}-3  & 0.40 \\ 
\ref{feat4} & \text{k}_\text{S/N} = \frac{\sum_\text{e} (\text{DM}_\text{e}-\overline{\text{DM}})^4\text{W}_\text{e}}{\sigma^4(\text{S/N}_\text{e})\sum_\text{e}\text{S/N}_\text{e}}-3 & 0.29 \\
\ref{feat5} & \text{S/N} = \max(\text{S/N}_\text{e}) & 0.10 \\
\bottomrule
\end{tabular}}
\end{table}

\subsubsection{Feature evaluation}
The classifier's ability to select astrophysical detections and to reject RFI instances is estimated using the manually-selected sample of pulses described in \textsection\ref{sec:training_set}.
Ideally, training and test sets should be composed of independent samples.
However, we chose to use the whole sample of astrophysical detections available to train the classifier.
Therefore, we used the same training set to test the classifier performance by running WEKA's ten fold cross validation, which randomly divides the sample into ten groups.
Nine of the groups are used to build a model and one to evaluate it. 
The process is iterated for each group and the average values are calculated.
Using the features and the ML algorithm described in the previous sections, $98.9\%$ of instances in the LOTAAS set are correctly classified and the resulting confusion matrix is reported in Table~\ref{tab:confusion}.

\begin{table}
\centering
\caption{Confusion matrix for the ML classifier of \textsc{L-SpS} applied to the LOTAAS set.}
\label{tab:confusion}
\begin{tabular}{lll}
%\toprule
 & Predicted & Predicted \\
 & pulsar & RFI \\
\cmidrule(lr){2-3}
Actual pulsar & TP = 17754 & FN = 249 \\
Actual RFI & FP = 355 & TN = 34708 \\
%\bottomrule
\end{tabular}
\end{table}

\begin{figure}
\centering
\includegraphics{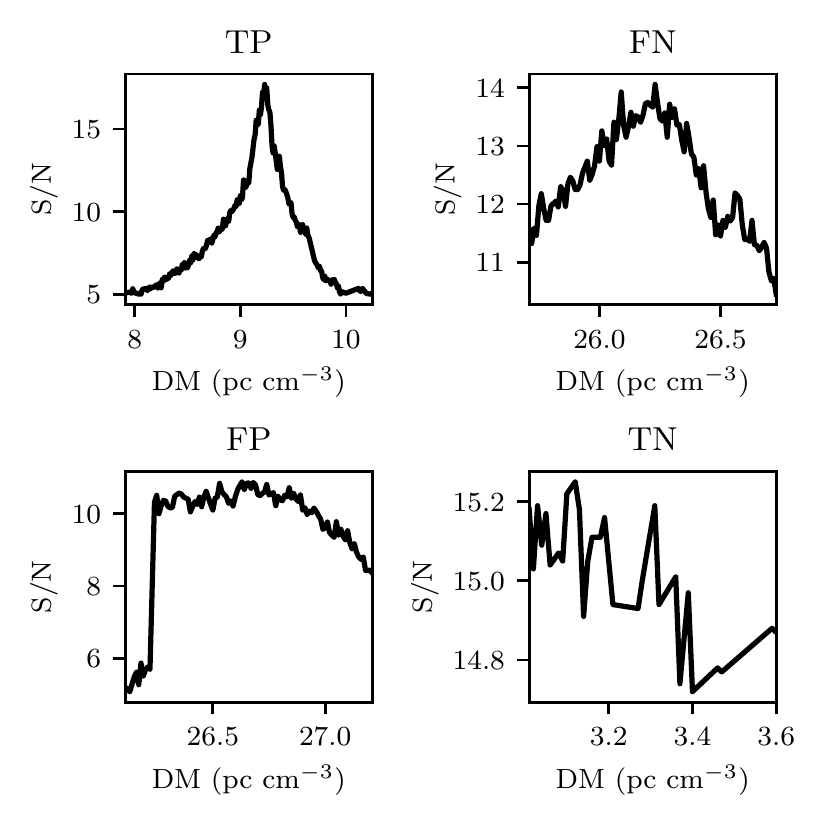}
\caption{S/N curves along DM of four pulses with the relative classification indicated on top.
}
\label{img/pulses_examples}
\end{figure}

A manual inspection of the FP and FN instances returned by the classifier on the training set was performed.
The vast majority of the mislabelled instances would have been misclassified even after visual inspection, with high probability.
It is possible that these instances were spurious detections incorrectly labelled when the training set was built (see \textsection\ref{sec:training_set}).
In rare cases, however, the instances misclassified by the ML classifier could be correctly labelled after visual inspection.
This is partially because astrophysical pulses affected by RFI sometimes mimic RFI behaviour and therefore some RFI is mislabelled in order to retain such astrophysical signals.
This could also be an indication that there is still space for additional features useful for discriminating RFI instances difficult to classify from true astrophysical signals.
The S/N curves along DM of four pulses are reported in Fig.~\ref{img/pulses_examples} as an example.

We have compared the performance of our classifier to others presented in the literature.
However, it is important to note that statistical metrics cannot be used to compare classifiers on different samples.
In fact, if we had chosen e.g.\ different thresholds on the pulses to build the training set (e.g.\ S/N or number of events) a different performance would have resulted.
Therefore, it is only for reference that we compare the performance of other single-pulse classifiers.
We only compare our ML classifier and not the full classification process (e.g.\ event grouping, beam comparison, etc.).
The values obtained for the different single-pulse classifiers are reported in Table~\ref{tab:classification}.

\begin{table*}
\centering
\caption{Metrics to evaluate the performance of different single-pulse classifiers.
The ML algorithm of \textsc{L-SpS} is applied to the LOTAAS training set.
The metrics used can assume any value between 0 and 1 and are described in the text.
While the FNR and FPR assume lower values for a better classification, the rest of the metrics assume higher values for a better classification.
\textsc{RRATtrap} and \textsc{Clusterrank} do not use ML and only rough estimations are available.
The metrics are evaluated on different datasets and thus they are reported only for reference (see main text for details).
Values for \textsc{RF$^2_\text{over}$} are from \citet{Dev16}, for \textsc{RRATtrap} are from \citet{Kar15} and for \textsc{Clusterrank} are from \citet{Den16}.
}
\label{tab:classification}
\begin{tabular}{lLLLLLLL}
\toprule
Classifier & \text{FNR} & \text{FPR} & \text{TPR} & \text{ACC} & \text{PPV} & \text{G-M} & \text{F-M} \\
\midrule
\textsc{L-SpS}              & 0.014 & 0.010 & 0.986 & 0.989 & 0.980 & 0.983 & 0.983 \\
\textsc{RF$^2_\text{over}$} & 0.282 & 0.011 & 0.718 & -     & -     & -     & 0.716 \\
\textsc{RRATtrap}     & 0.2 & 0.09 & -     & -     & -     & -     & -     \\
\textsc{Clusterrank}  & 0.3 & 0.07 & -     & -     & -     & -     & -     \\
\bottomrule
\end{tabular}
\end{table*} 

\subsection{Computational performance of the \textsc{L-SpS} classifier}\label{sec:performance}
% Number of instances from PRESTO to candidates
As mentioned in \textsection\ref{sec:observations}, the number of events that are stored at the end of the \textsc{PRESTO}-based pipeline is on the order of $10^7$-$10^8$ above $5 \sigma$ for a typical LOTAAS observation.
By only selecting one pulse for each group of events and by removing all the pulses formed by less than $6$ events, having $\text{S/N}<6.5$ or $\text{DM}<3$\,pc\,cm$^{-3}$, as discussed in \textsection\ref{sec:pulses}, the number of instances decreases to $\sim 10^{6}$.
At this point, pulses are selected using the ML classifier described in \textsection\ref{sec:ML}.
For a typical observation, the ML selection takes roughly a minute to process the whole observation.
At the end of the ML process $\sim10^4$ pulses remain.
The spatial comparison of different beams on the sky removes roughly half of these pulses.
The grouping of pulses into candidates, the selection of the brightest for each DM range and the thresholds on their cumulative S/N values discussed in \textsection\ref{sec:candidates} typically leaves $\sim20$ diagnostic files to inspect per observation.
In total, \textsc{L-SpS} takes around 30 minutes to process one observation on one Cartesius node (powered by $2\times12$ cores $2.6$\,GHz Intel Xeon E5-2690 v3 Haswell).
Multimoment analysis could further reduce the number of RFI instances and tests are ongoing. 

\begin{figure*}
\centering
\includegraphics{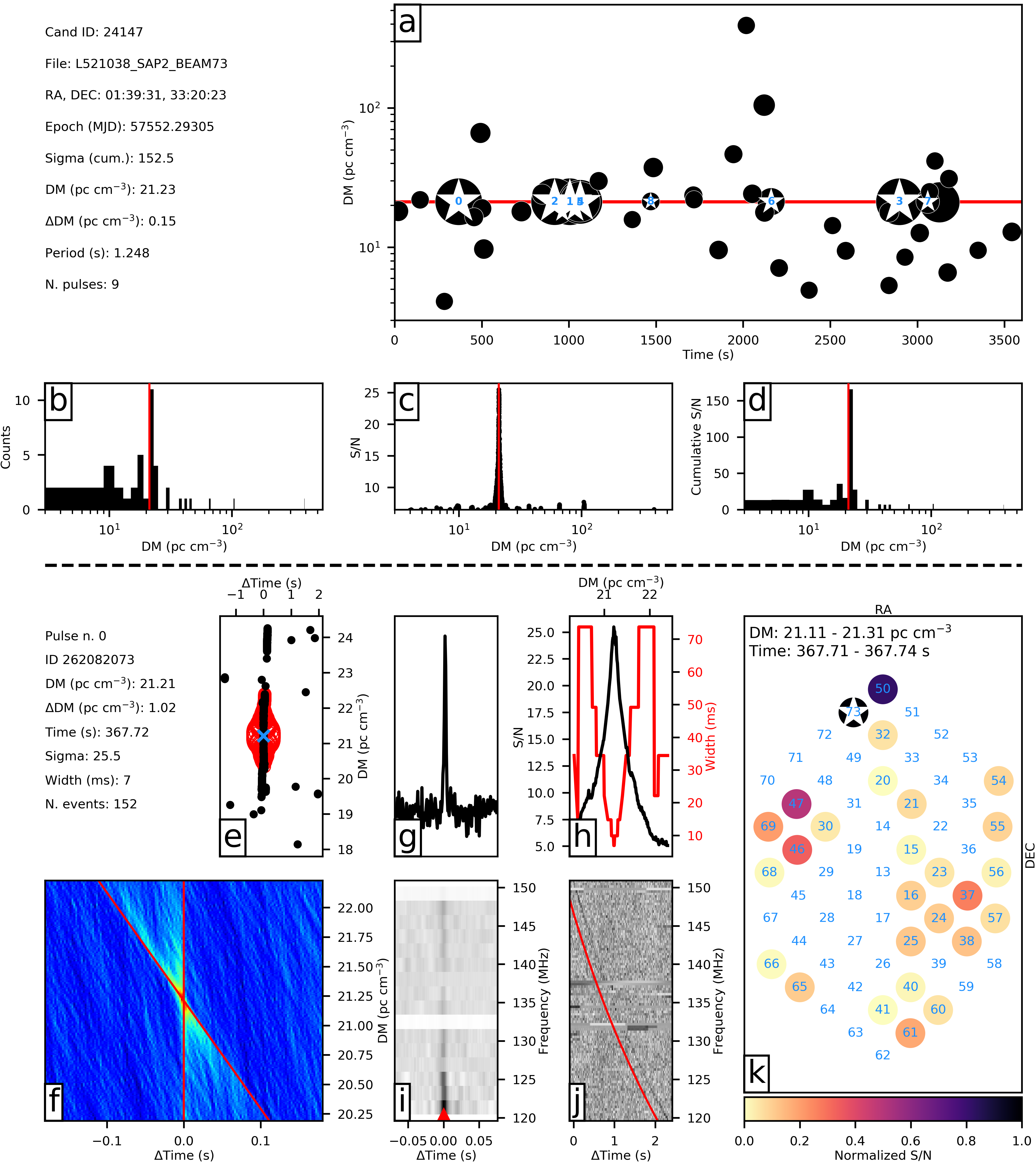}
\caption{Discovery plot of PSR J0139+33. Diagnostic plots like this are generated for each of the positively classified candidates from the LOTAAS survey.
 Top: summary plots of the pulses detected in the beam.
 Bottom: plots of the brightest pulse forming the candidate.
 The description of each sub-panel is given in \textsection\ref{sec:L-SpS:plot}.
}
\label{img/diagnostic_plot}
\end{figure*}

\subsection{Diagnostic plots}\label{sec:L-SpS:plot}
% Importance of diagnostic plots
Despite the huge progress in ML classification in recent years, a final human inspection is essential.
Bright signals can be easily identified by machines and by humans but for weak signals the classification becomes uncertain even after human inspection.
For this reason, we developed a set of diagnostic plots to help visually identify astrophysical signals even when they are buried in the noise.
Once an observation has been processed, a file containing the diagnostic plots is generated for each candidate.
An example is shown in Fig.~\ref{img/diagnostic_plot}, which represents the discovery plot of PSR~J0139+33.
For comparison, its detection in a confirmation observation is shown in Fig.~\ref{img/conf}, while a diagnostic plot for an RFI candidate is shown in Fig.~\ref{img/diagnostic_plot_RFI}.

% Beam page
In the top part, four summary plots of detections in the beam are present, along with meta data containing information on the beam where the candidate was detected. 
In Fig.~\ref{img/diagnostic_plot}a, all the pulses positively classified in the beam are plotted with a size proportional to their S/N as a function of time of arrival and DM. 
The pulses belonging to the candidate shown are highlighted with stars and the ten brightest are numbered in order of decreasing S/N. 
Fig.~\ref{img/diagnostic_plot}b reports the number of positively classified pulses per DM range. 
Fig.~\ref{img/diagnostic_plot}c shows the distribution of S/N vs DM of the events forming the pulses.
Fig.~\ref{img/diagnostic_plot}d shows the cumulative S/N of the pulses for each DM range.
The red lines indicate the candidate DM.

% Pulse pages
In the bottom part, the sub-panels show information related to the brightest pulse associated with the candidate source.
Meta data about the pulse is reported at the top-left. 
Fig.~\ref{img/diagnostic_plot}e is a close-up of the brightest pulse from Fig.~\ref{img/diagnostic_plot}a. 
It shows all the single events in the DM and time ranges plotted as black dots, including those considered to be noise or RFI.
Events forming the pulse are highlighted with red circles, where the size is proportional to the S/N.
The blue cross marks the most likely DM and time of the burst. 
Fig.~\ref{img/diagnostic_plot}f shows the time-series around the pulse time and DM.
Red lines highlight the expected smearing of a broad-band signal.
Fig.~\ref{img/diagnostic_plot}g shows the profile of the pulse, i.e. the time-series around the pulse's time of arrival de-dispersed at the pulse's DM.
Fig.~\ref{img/diagnostic_plot}h represents the S/N (black line) and width (red line) of the events forming the pulse as functions of DM.
Fig.~\ref{img/diagnostic_plot}i and j represent the pulse spectrum around the pulse's time of arrival.
These two plots are generated using \textsc{waterfaller.py}\fnurl{https://github.com/plazar/pypulsar}.
In Fig.~\ref{img/diagnostic_plot}i, the spectrum is de-dispersed at the pulse DM, down-sampled in frequency by a factor of 162 to 16 sub-bands and smoothed in time with a boxcar function of the same width as the pulse.
The red triangle indicates the expected position of the signal.
Fig.~\ref{img/diagnostic_plot}j shows the dispersed data around the pulse, down-sampled in frequency by a factor of 27 to 96 sub-bands and in time to 3 times the pulse width.
The red curve represents the expected DM sweep of the signal.
The RFI visible in Fig.~\ref{img/diagnostic_plot}i and j as horizontal stripes is the cause of the weak detections around the pulse in Fig.~\ref{img/diagnostic_plot}e and of the short-duration events in the pulse tails in Fig.~\ref{img/diagnostic_plot}h.
Fig.~\ref{img/diagnostic_plot}k reports the maximum S/N of the events detected in each of the coherent beams of the SAP, within the DM and time ranges indicated at the top.
The S/N is normalized between 0 and 1 and represented as a hot-shades colour-scale.
A white star indicates the beam where the candidate was detected and beams are numbered in blue.
In this case, the signal is strongest in beam 73, where the pulse is detected.
The complicated sensitivity pattern and RFI is responsible for the (apparent) weaker detections in other beams.

\section{Early single-pulse discoveries from the LOTAAS survey}\label{sec:results}
The \textsc{L-SpS} classifier has been developed and tested using data from the LOTAAS survey.  The latest version of the program is presented in this paper and it will be used in a complete reanalysis of the survey data.
Around 80 known pulsars have been blindly identified by the different preliminary versions of \textsc{L-SpS}.
Some of these were used to produce the training set described in \textsection\ref{sec:training_set}.
Seven new pulsars have been discovered to date using \textsc{L-SpS} --- dozens have also been discovered using the periodicity searches and associated classifier \citep{Tan18}.
A summary of their properties is reported in Table~\ref{tab:pulsars}.
PSR~J0139+33 is a RRAT that was not detectable in periodicity searches in the discovery pointing, nor in successive, targeted observations.
PSRs~J0301+20, J0317+13 and J1849+15 manifest a strong variability between single pulses.
For this reason, the first two were initially detected through their bright single pulses and subsequently also found using a periodicity search in follow-up observations.
The remaining pulsars are bright enough to be detected in both periodicity and single-pulse searches.
The timing solutions of the sources presented here will be reported in future papers.

\begin{table}
\centering
\caption{Pulsars discovered by early versions of \textsc{L-SpS} in the LOTAAS survey.
The level of pulse brightness variability during the observation is described qualitatively.}
\label{tab:pulsars}
\begin{tabular}{lCCl}
\toprule
Name & \text{Period (s)} & \text{DM (pc\,cm$^{-3}$)} & Variability \\
\midrule
PSR~J0139+33   & 1.248 & 21.2 & Extreme \\
PSR~J0301+20   & 1.207 & 19.0 & Large \\
PSR~J0317+13   & 1.974 & 12.9 & Large \\
PSR~J0454+45   & 1.389 & 20.8 & Some \\
PSR~J1340+65   & 1.394 & 30.0 & Some \\
PSR~J1404+11   & 2.650 & 18.5 & Some \\
PSR~J1849+15   & 2.233 & 77.4 & Extreme \\
\bottomrule
\end{tabular}
\end{table}

\section{Conclusions and future developments}\label{sec:conclusions}
We have presented \textsc{L-SpS}, a new classifier for searches of single radio pulses in the LOTAAS survey.
It employs a ML algorithm to discriminate astrophysical signals from RFI, with high accuracy.
During its development, the algorithm has discovered 7 new pulsars and blindly identified $\sim$80 known sources.
A full reprocessing of the LOTAAS data with the latest version of \textsc{L-SpS}, as presented here, is under way.

Future improvements to the software include testing of multimoment analysis and development of additional features.
Also, we only made use of binary classification, i.e. instances were divided into astrophysical and RFI.
The use of multiclass classification (e.g.\ distinguishing broad-band and narrow-band RFI) could improve the performance \citep{Tan18}.
In addition, a larger training set with positive instances better distributed (e.g.\ more high-DM pulsars) will be used in future.
Finally, deep learning techniques could significantly improve the classification performance \citep{Den14}.
However, deep learning algorithms typically require larger training sets.
Therefore, they could possibly be used to reprocess the survey data when a sufficient number of discoveries and re-detections is achieved.

\subsection{Portability}
Although \textsc{L-SpS} has been designed specifically for the LOTAAS survey, efforts are ongoing to produce a more general software \citep[\textsc{SpS},][]{mic18} that can be readily adapted to other projects.
The aim is to create a program that is user-friendly, simple to customize and robust to different observation characteristics, in order to be easily used in a generic study.
This is achieved by designing a modular software where the different tasks discussed in \textsection\ref{sec:L-SpS} are performed by different modules executed in sequence by a main script.
Therefore, the sequence can be easily modified and each of the modules can be tailored to a specific study.
A first release of this software is freely available on github\fnurl{https://github.com/danielemichilli/SpS}.
At the time of writing, while the correct operation of \textsc{SpS} has been extensively tested, some of the features developed specifically for the LOTAAS survey still need to be included, such as the capability to process multiple telescope beams in parallel over different computer cores. 

Arguably the most critical part of the program is the final selection of astrophysical signals.
In fact, due to the need for a large data set of labelled detections, an ML analysis can be difficult or impossible to perform in the case of small studies.
Therefore, a set of filters that do not rely on ML techniques was created for these situations.
While such filters have been designed to keep the false negative rate low, the false positive rate will be higher than in the ML approach, though still manageable for small projects.
It is difficult to assess the general performance of these filters since they depend on the characteristics of the specific observations.
A rough estimate is obtained from the study of FRB~121102 with Arecibo.
The features used typically reduce the number of candidates by $\sim 80$-$90$\%.
Of the remaining candidates, typically $\sim$25\% were found to be real after visual inspection, though this fraction varied between 1\% and 64\% in different observations (depending on the severity of RFI). 

\section*{Acknowledgements} 
This work was carried out on the Dutch national e-infrastructure with the support of SURF Cooperative. Computing time was provided by NWO Physical Sciences (project SH-242-15).
The LOFAR facilities in the Netherlands and other countries, under different ownership, are operated through the International LOFAR Telescope foundation (ILT) as an international observatory open to the global astronomical community under a joint scientific policy.
DM and JWTH acknowledge funding from the European Research Council under the European Union's Seventh Framework Programme (FP/2007-2013) / ERC Starting Grant agreement nr. 337062 (``DRAGNET'').
JWTH also acknowledges funding from an NWO Vidi fellowship.
JvL acknowledges funding from the European Research Council under the European Union's Seventh Framework Programme (FP/2007-2013) / ERC Grant Agreement n. 617199.
Different Python modules were used in this study, specifically \textsc{numpy}, \textsc{matplotlib}, \textsc{pandas} and \textsc{astropy}.

\bibliographystyle{mnras}
\bibliography{michilli}

\appendix
\section{RFI diagnostic plot}
\begin{figure*}
\centering
\includegraphics{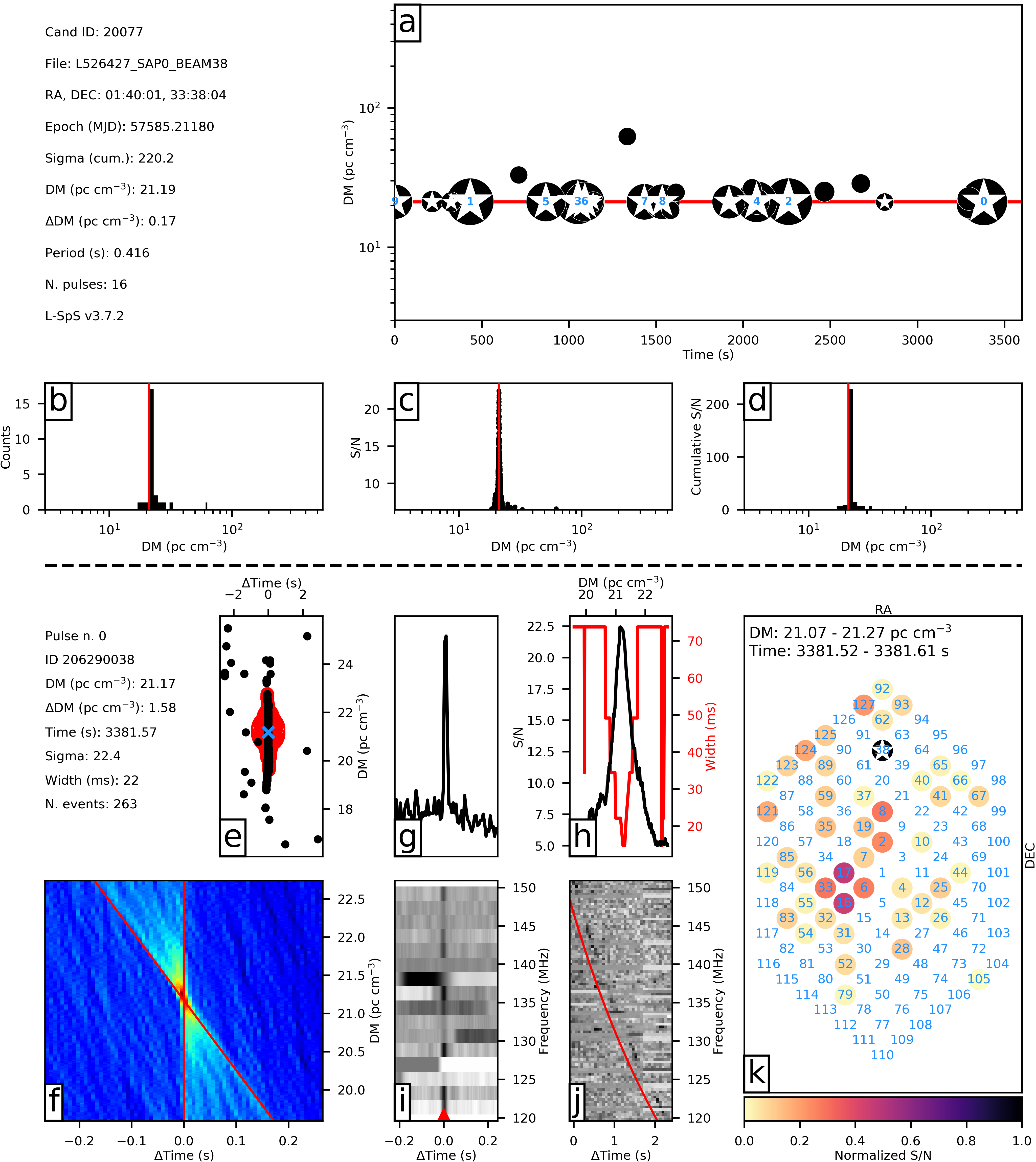}
\caption{Example of diagnostic plots of the confirmation observation of PSR~J0139+33. The same plots as Fig.~\ref{img/diagnostic_plot} are represented and they are discussed in the text.
The number of beams visible in panel k is different for confirmation observations.
The strong interference visible as horizontal stripes in panels i and j did not prevent to detect the bright pulses from the RRAT.}
\label{img/conf}
\end{figure*}

\begin{figure*}
\centering
\includegraphics{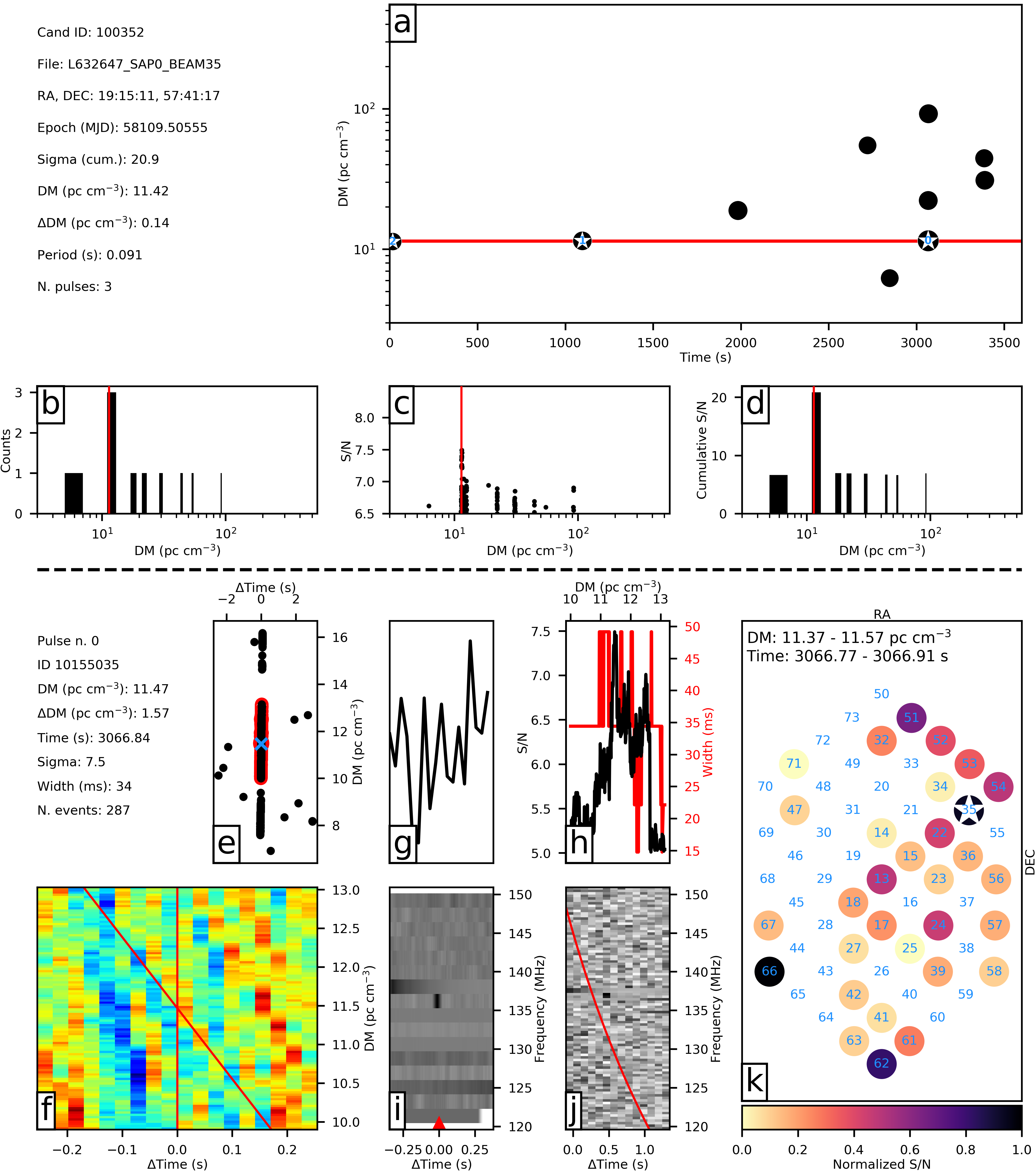}
\caption{Example of diagnostic plots for an RFI candidate. The same plots as in Fig.~\ref{img/diagnostic_plot} are represented and they are discussed in the text.
}
\label{img/diagnostic_plot_RFI}
\end{figure*}

Examples of two diagnostic plots generated by \textsc{L-SpS} are presented.
Fig.~\ref{img/conf} reports the detection of PSR~J0139+33 during its confirmation observation.
Fig.~\ref{img/diagnostic_plot_RFI} shows an example of a typical diagnostic plot for a candidate produced by RFI.
The explanation of the sub-plots is given in \textsection\ref{sec:L-SpS:plot}.

\bsp	% typesetting comment
\label{lastpage}
\end{document}